# Comments on "Prediction of Subharmonic Oscillation in Switching Converters Under Different Control Strategies"

## Chung-Chieh Fang

*Abstract*—A recent paper [1] (El Aroudi, 2012) misapplied a critical condition (Fang and Abed, 2001) to a well-known example. Even if the mistake is corrected, the results in [1] are applicable only to buck converters and period-doubling bifurcation. Actually, these results are known in Fang's works a decade ago which have *broader* critical conditions applicable to other converters and bifurcations. The flaws in [1] are identified.

## I. INTRODUCTION

Recently the author of [1], based on [2], presents critical conditions for current/voltage mode control (CMC/VMC) for *buck* converters. The applicability of critical conditions is not stated, which may mislead or confuse the readers. As an author of [2], I feel a need to clarify the results in [1]. Actually, the results are known in [3], [4], [5], [6], [7], [8], [9] a decade ago which have *broader* critical conditions applicable to other converters and bifurcations. The author of [1] was aware of these results when he received an email copy of [3] on Aug. 6, 2010 with the key results *particularly identified*, but [3] was not cited in [1].

## II. BROADER RESULTS KNOWN A DECADE AGO

### A. Unified VMC/CMC Model With a Common Ramp

In VMC, a sawtooth signal is used for PWM *modulation*. In CMC, a ramp signal is used for *stabilization*. The two signals serve different purposes. However, they have the same waveform. In [3], [4], [5], [6], they are modeled by the same signal $h(t)$ in a unified VMC/CMC model shown in Fig. 1, where $h(t) := V_l + (V_h - V_l)(\frac{t}{T} \bmod T)$ varies from a low value $V_l$ to a high value $V_h$ with an amplitude of $V_m = V_h - V_l$.

Since the model is applicable to both VMC and CMC, the system dynamics and critical (boundary) conditions associated with the unified model are also applicable to both VMC and CMC. No special or separate analysis is required, and the unified model has broad applications.

Also, the unified model is applicable to both the power stage and the closed-loop converter (power stage plus compensator). For the power stage, a *control* signal is used to control the inductor current $i_L$ in CMC or the output voltage $v_o$ in VMC.

C.-C Fang is with Advanced Analog Technology, 2F, No. 17, Industry E. 2nd Rd., Hsinchu 300, Taiwan, Tel: +886-3-5633125 ext 3612, Email: fangcc3@yahoo.com

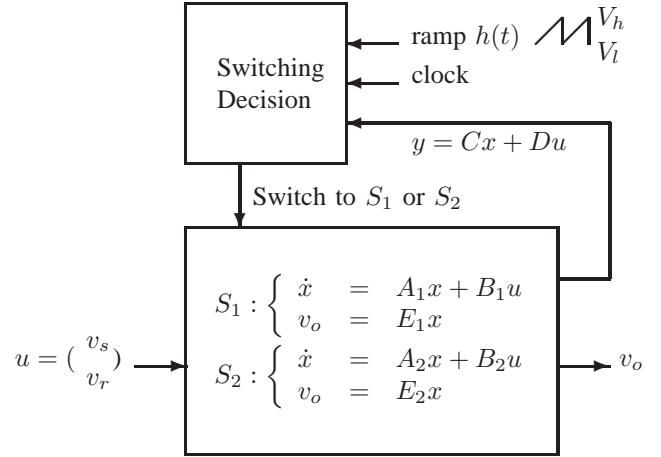

Figure 1. Unified VMC/CMC model for a switching converter.

For the closed-loop converter, a *reference* signal is used for similar purposes. The two signals are modeled as $v_r$. Let the source voltage be $v_s$. Both $v_r$ and $v_s$ are external signals input into the converter, collectively represented by a vector $u$.

In the model, $A_1, A_2 \in \mathbf{R}^{N \times N}$, $B_1, B_2 \in \mathbf{R}^{N \times 2}$, $C, E_1, E_2 \in \mathbf{R}^{1 \times N}$, and $D \in \mathbf{R}^{1 \times 2}$ are constant matrices, where $N$ is the system dimension. Within a clock period $T$, the dynamics is switched between two stages, $S_1$ and $S_2$. Let the compensator output be $y := Cx + Du \in \mathbf{R}$. The dynamics is switched to $S_1$ at $t = nT$. The dynamics is switched to $S_2$ when $h(t) \geq y(t)$. In the trailing-edge modulation (TEM), $S_1$ is the ON stage, and $S_2$ is the OFF stage. In the leading-edge modulation (LEM), $S_1$ is the OFF stage, and $S_2$ is the ON stage. Denote the switching frequency as $f_s := 1/T$ and let $\omega_s := 2\pi f_s$.

### B. Closed-Form Jacobian Matrix, First Known in [3], [4]

The periodic solution $x^0(t)$ of the system in Fig. 1 corresponds to a fixed point $x^0(0)$ in the sampled-data dynamics. Let $\dot{x}^0(d^-) = A_1 x^0(d) + B_1 u$ and $\dot{x}^0(d^+) = A_2 x^0(d) + B_2 u$ denote the time derivative of $x^0(t)$ at $t = d^-$ and $d^+$, respectively. Let $y^0(t) = Cx^0(t) + Du$. In steady state, $\dot{y}^0(t) = C\dot{x}^0(t)$. Let the steady-state duty cycle be $D$ and let $d$ be the switching instant within a cycle. For LEM, $d = (1-D)T$; for TEM, $d = DT$. Confusion of notations for capacitance $C$ and duty cycle $D$ with the matrices $C$ and $D$ can be avoided from the context.

In steady state,

$$x^0(d) = e^{A_1 d} x^0(0) + \int_0^d e^{A_1 \sigma} d\sigma B_1 u \quad (1)$$

$$x^0(0) = e^{A_2(T-d)} x^0(d) + \int_0^{T-d} e^{A_2 \sigma} d\sigma B_2 u \quad (2)$$

$$y^0(d) = C x^0(d) + D u = h(d) \quad (3)$$

Solving (1)-(3), one obtains $d$ and $x^0(d)$.

Using a hat $\hat{\ }$ to denote small perturbations (e.g., $\hat{x}_n = x_n - x^0(0)$), where $x_n$ is the sampled state at $t = nT$. From [3], [4], [5], the linearized sampled-data dynamics is

$$\hat{x}_{n+1} = \Phi \hat{x}_n \quad (4)$$

where the Jacobian matrix, *first known* in [3], [4], is

$$\Phi = e^{A_2(T-d)}(I - \frac{(\dot{x}^0(d^-) - \dot{x}^0(d^+))C}{C\dot{x}^0(d^-) - \dot{h}(d)})e^{A_1 d} \quad (5)$$

A closed form like (5) is important, because it leads to many general critical conditions discussed next.

### C. Equivalent Linearized Sampled-Data Dynamics

Let $\Phi = \Phi_0 - \Gamma \Psi$, where

$$\Phi_0 = e^{A_2(T-d)} e^{A_1 d} \quad (6)$$

$$\Gamma = e^{A_2(T-d)}(\dot{x}^0(d^-) - \dot{x}^0(d^+)) \quad (7)$$

$$\Psi = \frac{C e^{A_1 d}}{C\dot{x}^0(d^-) - \dot{h}(d)} \quad (8)$$

The dynamics (4) can be transformed into a plant (having an input $\hat{d}_n$ and an output $\hat{w}_n = \Psi \hat{x}_n$) with a *unity negative* feedback ($\hat{d}_n = -\hat{w}_n$).

$$\begin{aligned} \hat{x}_{n+1} &= \Phi_0 \hat{x}_n + \Gamma \hat{d}_n \\ \hat{d}_n &= -\Psi \hat{x}_n \end{aligned} \quad (9)$$

### D. Broader General Closed-Form Critical Conditions

**Theorem 1:** [3, p. 46] Suppose that $\lambda$ is not an eigenvalue of $\Phi_0$. Then $\lambda$ is an eigenvalue of $\Phi$ if and only if

$$C\dot{x}^0(d^-) + C e^{A_1 d}(\lambda I - \Phi_0)^{-1} \Gamma = \dot{h}(d) \quad (10)$$

*Proof:* Suppose $\lambda$ is not an eigenvalue of $\Phi_0$.

$$\begin{aligned} \det[\lambda I - \Phi] &= \det[\lambda I - \Phi_0] \det[I + (\lambda I - \Phi_0)^{-1} \Gamma \Psi] \\ &= \det[\lambda I - \Phi_0](1 + \Psi(\lambda I - \Phi_0)^{-1} \Gamma) \end{aligned}$$

Then $\det[\lambda I - \Phi] = 0$ requires that $\Psi(\lambda I - \Phi_0)^{-1} \Gamma = -1$, which leads to (10) by using (6)-(8). $\square$

Instability occurs when there exists an eigenvalue $\lambda$ (also the sampled-data pole) of $\Phi$ outside the unit circle of the complex plane, which leads to the following corollary, associated with three typical instabilities [3], [7], [9]: period-doubling bifurcation (PDB, subharmonic oscillation), saddle-node bifurcation (SNB), and Neimark-Sacker bifurcation (NSB).

**Corollary 1:** [3, p. 46] (i) If the system parameters correspond to an occurrence of PDB ($\lambda = -1$), then

$$C\dot{x}^0(d^-) - C e^{A_1 d}(I + \Phi_0)^{-1} \Gamma = \dot{h}(d) \quad (11)$$

(ii) If the system parameters correspond to an occurrence of SNB ($\lambda = 1$), then

$$C\dot{x}^0(d^-) + C e^{A_1 d}(I - \Phi_0)^{-1} \Gamma = \dot{h}(d) \quad (12)$$

(iii) If the system parameters correspond to an occurrence of NSB ($\lambda = e^{j\theta}$, $\theta \neq 0$ or $\pi$), then

$$C\dot{x}^0(d^-) + C e^{A_1 d}(e^{j\theta} I - \Phi_0)^{-1} \Gamma = \dot{h}(d) \quad (13)$$

All of the results above are applicable to *general* converters, they are applied to a special case: *buck* converters.

### E. Applying General PDB Condition (11) to Buck Converter

Let $B_1 := [B_{11}, B_{12}]$, $B_2 := [B_{21}, B_{22}]$ to expand the matrices into two columns. Many LEM and TEM examples are presented in [3].

The LEM buck converter generally has

$$A_1 = A_2 = A, \quad B_1 = [0_{N \times 1}, B_{12}], \quad B_2 = [B, B_{12}] \quad (14)$$

Using (1) and (2), the PDB critical condition (11) becomes

$$C[(I + e^{-AT})^{-1} + (I - e^{-AT})^{-1}(e^{AT} - e^{Ad})]B v_s = \dot{h}(d) \quad (15)$$

or in terms of $v_s$, shown in [3, Eq. 3.118], with $d = (1-D)T$,

$$v_s^{\text{LEM}}(D) = \frac{\dot{h}(d)}{C[(I + e^{-AT})^{-1} + (I - e^{-AT})^{-1}(e^{AT} - e^{AT(1-D)})]B} \quad (16)$$

### F. Directly Applying (11) to TEM: Different Dynamics

The TEM buck converter generally has

$$A_1 = A_2 = A, \quad B_1 = [B, B_{12}], \quad B_2 = [0_{N \times 1}, B_{12}] \quad (17)$$

Using (1) and (2), the PDB critical condition (11) becomes

$$C[(I - e^{AT})^{-1}(e^{Ad} - I) + (I + e^{AT})^{-1}]B v_s = \dot{h}(d) \quad (18)$$

or in terms of $v_s$, with $d = DT$,

$$v_s^{\text{TEM}}(D) = \frac{\dot{h}(d)}{C[(I - e^{AT})^{-1}(e^{ATD} - I) + (I + e^{AT})^{-1}]B} \quad (19)$$

It is known decades ago that TEM and LEM have different dynamics [10], [11]. Thus, $v_s^{\text{TEM}}(D) \neq v_s^{\text{LEM}}(D)$. However, [1] wrongly assumes $v_s^{\text{TEM}}(D) = v_s^{\text{LEM}}(D)$. As a side note, the recent results in [12], [13] were *repeatedly* said to be wrong by a reviewer on this issue. That reviewer was wrong. One can prove $v_s^{\text{TEM}}(D) = -v_s^{\text{LEM}}(1 - D)$ by simple algebra.

### G. Equivalent Critical Condition Based on Harmonic Balance

Let $\mathbf{Re}$ denote taking the real part of a complex number. Let the voltage across the second switch (or the diode) be $v_d$. Let the $v_d$-to-$y$ transfer function be $G(s)$ for LEM, or $-G(s)$ for TEM. Based on harmonic balance analysis, an *equivalent* PDB critical condition for the buck converter (due to its special property (14) or (17)) is obtained [2], [3], [14],

$$v_s = \frac{V_m}{2\mathbf{Re}\left[\sum_{k=1}^{\infty}[(1-e^{j2k\omega_s d})G(jk\omega_s)-G(j(k-\frac{1}{2})\omega_s)]\right]}$$
(20)

As reported in [3, pp. 72-73], both (20) and (16) are *exact* PDB conditions, and they are the same. Therefore,

$$2f_s\mathbf{Re}\left[\sum_{k=1}^{\infty}[(1-e^{j2k\omega_s d})G(jk\omega_s)-G(j(k-\frac{1}{2})\omega_s)]\right]$$
$$= C[(I+e^{-AT})^{-1}+(I-e^{AT})^{-1}(e^{AT}-e^{Ad})]B \quad (21)$$

Note that (15), (18), (20), and (21), expressed in terms of $d$, are valid for *both* LEM and TEM ($d = (1 - D)T$ for LEM, $d = DT$ for TEM), whereas the main result of [1], (17) (underline added for [1]), is limited to TEM only.

### III. Simple Direct Extension of (18)

Using Taylor series expansion, (18) *directly* leads to, also shown in [12] submitted seven months earlier than [1],

$$C(\sum_{n=0}^{\infty}\delta_n(D)A^nT^n)Bv_s = \dot{h}(d) \quad (22)$$

where $\delta_0(D) = (1 - 2D)/2$, $\delta_1(D) = (-1 + 2D - 2D^2)/4$, and $\delta_2(D) = (-D + 3D^2 - 2D^3)/12$. If the value of a real pole (eigenvalue of $A$) is comparable with $\omega_s$, the high order terms in (22) are *significant* and cannot be ignored [12].

Note: (17) of [1] is exactly (19), and (20) is exactly (22).

### IV. Flaws, Mistakes, and Confusions in [1]

First, all of the obtained results in [1] are applicable to *buck* converters only, and therefore the title of [1] without this limitation is misleading.

Second, the condition (20) is applied without stating its applicability. In (20), $d = (1 - D)T$ for LEM. However, $d = DT$ is used instead in [1]. The mistake occurs at the *beginning* and all the results in [1] depend on proper application of (20).

Third, throughout [1], different TEM and LEM examples are commingled. It is probably assumed in [1] that any critical condition is *applicable to both*, which is wrong. If the goal of [1] is on the TEM conditions, then a wrong TEM condition is misapplied to the *well-known LEM* example on page 5. If the goal is on the LEM conditions, in contrary, all results (5)-(24) of [1] are valid only for TEM. The *sequence* of presentation in [1] is confusing: an LEM condition is misapplied with $d = DT$ (actually for *TEM*), then some TEM conditions are derived but applied to an *LEM* example on page 5.

Fourth, readers are misled about the cited references. On page 2, it is stated, "in [14], the series $\mathcal{S}(D)$ in (5) has been approximated by the term that involves the transfer function $H(s)$ with the smallest argument or graphically solved by truncating the series to an arbitrarily large number of harmonics. In this work, a closed form expression will be given for the series $\mathcal{S}(D)$ defined in (5)." It sounds like a different approach is used in [1]. Actually, still a *series* closed form (20) is presented, and eventually still a *figure* (Fig. 3) is used to obtain the critical value. In [14], the approximation is just one way to obtain the critical value. The key in [14] is on the *exact* condition (5), on which *all* of the results of [1] are based.

Fifth, great efforts (from (5) to (15)) are made in [1] to paraphrase the *equality* (21) *already* reported in [3, pp. 72-73]. As discussed above, (15) for the buck converter is just a *special* case of the *known* condition (10). Note that the condition (10) is not only broader but its derivation is also simpler, without the need of the great efforts in [1].

Sixth, the author of [1] is aware of the *general* condition (10) available in [3] and the *unified* VMC/CMC model [3], [4], [5], [6], [7], [8] of Fig. 1. However, none of [3], [4], [5], [6], [7], [8] is cited. Before 1997, to my knowledge, the sawtooth signal in VMC and the stabilization ramp in CMC were not modeled by a *single* signal.

Seventh, some parts are inconsistent with the others in [1]. Using (21) does not lead to Table I in [1], unless (21) is scaled by $1/(2\pi)^m$. Even with this corrected, using Table I does not lead to (4), unless $\mathcal{S}_1(D) = D - 1/2$ is corrected in Table I.

Eighth, the TEM condition (17) is misapplied to the *well-known LEM* example on page 5.

Ninth, the readers are not able to *reproduce* the same results. In Fig. 3 of [1], a plot is falsely *claimed* to be based on (17), but actually based on (16) *already known* in [3]. As stated in [1], $g\mathbf{C} = (8.4, 0)$ for this example. The plot of (17) is actually all *negative* (not shown). If it is corrected with $g\mathbf{C} = (-8.4, 0)$, the plot of (17) as shown in Fig. 2 is still incorrect because wrong $v_s^* = 35$ is obtained, not $v_s^* = 31$ as claimed in [1]. As expected, (17) is for TEM, not applicable to this LEM example. The correct (16) for LEM (*known* in [3] but never obtained in [1]) shown in Fig. 2 produces the correct $v_s^* = 31$. As expected, due to $v_s^{\text{TEM}}(D) = -v_s^{\text{LEM}}(1 - D)$, the correct plot is symmetric to the wrong one with respect to $D = 1/2$. The plot in Fig. 3 is not based on (17) as claimed. *How* the plot was produced to obtain $v_s^* = 31$ is a mystery.

Tenth, no experimental prototype is actually shown.

### V. Conclusion and Recent Developments

Compared with [1], broader results [3], [4], [5], [6], [7], [8], [9] known a decade ago are presented in this note. Actually, [1] is inconsistent with the past works of its own author. In the last six years, the author of [1] advocated a ripple index [15] to predict PDB for VMC. As indicated in [1], the author of [1] still up to *now* does not *distinguish* TEM from LEM (*known* with different dynamics), this gives doubt to the validity of his similar past works, further indicating that the ripple index has *limited* applicability. Actually using (5), which [1] totally relies on, *many* counter-examples of the ripple index hypothesis can be obtained [13], [16]. In fact the author of [1] attended the *same conference session* a decade ago when a counter-example (for average current control *known* to resemble VMC [17]) was presented [18]. The author of [1] was fully aware of this counter-example, never cited [18], and still advocated the ripple index in the last six years.

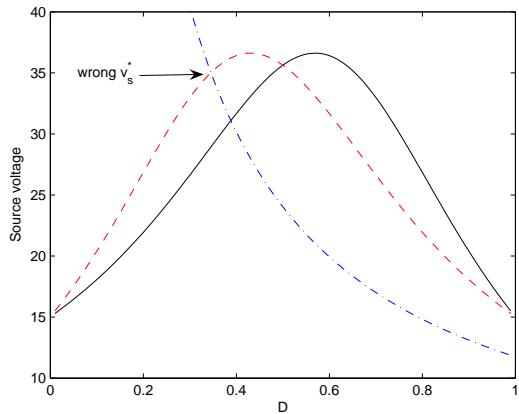

Figure 2. Dashed line based on the *claimed* (17) which leads to wrong $v_s^*$; solid line based on the correct (16) known 15 years ago [3, Eq. 3.118].

The readers are advised to refer to some *direct* extensions [12], [13], [16], [19], [20], [21] based on the key general condition (10). Let the left side of (10) be called an "S-plot" $S(\lambda, D)$, a function of $\lambda$ and $D$, for example. It predicts the occurrence of *three* typical instabilities (PDB, SNB, NSB) in DC-DC converters and shows the required ramp *slope* for stabilization. Let the loop gain transfer function of (9) be $\mathcal{N}(z)$. One has $\mathcal{N}(z) = \Psi(zI - \Phi_0)^{-1}\Gamma_0$. The (discrete-time) Nyquist plot is $\mathcal{N}(e^{j\omega T})$ as a function of $\omega$ swept from 0 to $\omega_s$. Also, define an "F-plot" as $F(\theta) = S(e^{i\theta}, D)$ swept from $-\pi$ to $\pi$. In a *single* plot in the complex plane, either Nyquist plot or F-plot also predicts the *three* typical instabilities.

The *broader* general critical conditions (10)-(13) have not been published except in my thesis, available online since 1997. These important and subsequent results [12], [13], [16], [19], [20], [21], *much broader than [1] and applicable to other converters and bifurcations, and variable switching frequency case* were fiercely opposed after submitted by a reviewer who was actually wrong about LEM/TEM. Only a small subset of key results (resubmitted to other journals) is about to be published [22], [23]. Many key results based on (10)-(13), although available since 1997, remain unpublished.